\def\simlt{\mathrel{\spose{\lower 3pt\hbox{$\mathchar''218$}}
     \raise 2.0pt\hbox{$\mathchar''13C$}}}
\def\simgt{\mathrel{\spose{\lower 3pt\hbox{$\mathchar''218$}}
     \raise 2.0pt\hbox{$\mathchar''13E$}}}
\documentclass[]{article}
\usepackage{emulateapj}
\usepackage{epsfig}

\begin{document}
\def\gtorder{\mathrel{\raise.3ex\hbox{$>$}\mkern-14mu
             \lower0.6ex\hbox{$\sim$}}}
\def\ltorder{\mathrel{\raise.3ex\hbox{$<$}\mkern-14mu
             \lower0.6ex\hbox{$\sim$}}}

\def\today{\number\year\space \ifcase\month\or  January\or February\or
        March\or April\or May\or June\or July\or August\or
        September\or
        October\or November\or December\fi\space \number\day}
\def\fraction#1/#2{\leavevmode\kern.1em
 \raise.5ex\hbox{\the\scriptfont0 #1}\kern-.1em
 /\kern-.15em\lower.25ex\hbox{\the\scriptfont0 #2}}
\def\spose#1{\hbox to 0pt{#1\hss}}
\def\rsun{$R_{\odot}$}
\def\msun{$M_{\odot}$}
\def\rstar{$R_{star}$}
\def\rplanet{$R_{p}$}
\title{Constraining the Rotation Rate of Transiting
Extrasolar Planets by Oblateness Measurements}
\author{S. Seager\footnote{Institute for Advanced Study, Einstein Drive, Princeton, NJ 08540; {\tt seager@ias.edu}}
and Lam Hui\footnote{Department of Physics,
Columbia University, 538 West 120th Street, New York, NY 10027; {\tt lhui@astro.columbia.edu}}}

\pagestyle{plain}

\begin{abstract}

The solar system gas giant planets are oblate due to their rapid
rotation.  A measurement of the planet's projected oblateness would
constrain the planet's rotational period. Planets that are
synchronously rotating with their orbital revolution will be rotating
too slowly to be significantly oblate; these include planets with
orbital semi-major axes $\lesssim $ 0.2~AU (for $M_P \sim M_J$ and
$M_* \sim M_{\odot}$).  Jupiter-like planets in the range of orbital
semi-major axis 0.1~AU to 0.2~AU will tidally evolve to synchronous
rotation on a timescale similar to main sequence stars' lifetimes. In
this case an oblateness detection will help constrain the planet's
tidal $Q$ value.

The projected oblateness of a transiting extrasolar giant planet is
measurable from a very high-photometric-precision transit light curve.
For a sun-sized star and a Jupiter-sized planet the normalized flux
difference in the transit ingress/egress light curve between a
spherical and an oblate planet is a few to 15$\times 10^{-5}$ for
oblateness similar to Jupiter and Saturn respectively.  The transit
ingress and egress are asymmetric for an oblate planet with an
orbital inclination different from 90$^{\circ}$
and a non-zero projected obliquity.  A photometric
precision of $10^{-4}$ has been reached by HST observations of the
known transiting extrasolar planet HD~209458~b.
Kepler, a NASA discovery-class mission designed to detect transiting
Earth-sized planets requires a photometric precision of $10^{-5}$ and
expects to find 30 to 40 transiting giant planets with orbital
semi-major axes $< 1$~AU, about 20 of which will be at $> 0.2$~AU.
Furthermore, part-per-million photometric precision (reached after
averaging over several orbital periods) is expected from three other
space telescopes to be launched within the next three years. Thus an
oblateness measurement of a transiting giant planet is realistic in
the near future.

\end{abstract}
Subject headings: planetary systems --- planets and satellites: general

\section{Introduction}
Extrasolar transiting planets are suitable for a variety of followup
measurements while they transit their parent star. Many of the
followup measurements would aim to detect additional planet components
that block light from the parent star, such as moons, rings, or the
planet's atmosphere. Recently, the atmosphere of the transiting planet
HD~209458~b was detected in transmission in the neutral sodium
resonance line (Charbonneau et al. 2002).  A followup measurement of a
transiting planet's albedo and phase curve would also be very useful
in constraining atmosphere properties since both the radius and
orbital inclination ($i$) are known (Seager, Whitney, \& Sasselov
2000). Here we describe an additional planet property that can be
derived from followup measurements of transiting extrasolar planets:
the planet's projected oblateness. The oblateness is defined as
$(R_e-R_p)/R_p$ where $R_e$ is the planet's equatorial radius and
$R_p$ the planet's polar radius.  The solar system gas giants all have
a significant oblateness, ranging from 0.02 to 0.1, due to their rapid
rotation.  Saturn's oblateness is so large it is obvious from
whole-planet images.  An oblateness detection would immediately tell
us that a planet is a rapid rotator.

A planet's oblateness signature on a transit light curve was first
described in Hui \& Seager (2002). The oblateness signature is
expected to be small. However, recent observations by HST/STIS (Brown
et al. 2001) reached an unprecedented photometric precision of
$10^{-4}$, and space telescopes MOST (Canadian Space Agency; launch
date December 2002), MONS (Danish Space Agency; launch date $\sim$
2004), Corot (French Space Agency, European Space Agency, launch date
2004), and Kepler (NASA; launch date 2006) are designed to reach
photometric precision of a few$\times 10^{-6}$ by averaging
measurements over several orbital periods. Kepler, a wide-field
transit search with the main goal of detecting Earth-sized planets
orbiting solar type stars, expects to find 30--40 transiting extrasolar
giant planets. With the prospect of high photometric precision, we
investigate the transit signature of an oblate giant planet and the
constraints on an oblate extrasolar planet's rotational period and
tidal $Q$ value.

\section{Oblateness and Rotational Period}
\label{oblatenessperiod}
Here we review the relation between oblateness and rotational
period (see e.g. Collins 1963; van Belle et al. 2001), making the
simplest assumptions about the structure of the planets in question.
The potential for a uniformly rotating body has two terms,
gravitational and rotational.  The gravitational potential for a
rotating, not too aspherical, figure of equilibrium can be expanded
using the Legendre polynomials (see e.g. Danby 1962; Chandrasekhar
1969),
\begin{equation}
\label{eq:gravpot}
\phi_g = -\frac{GM_P}{R}(1 - \frac{J_2}{R^2}\frac{1}{2}(3 \cos^2\theta - 1)
- \ldots ),
\end{equation}
where $G$ is the Universal gravitational constant, $M_P$ is the planet
mass, $R$ is the planet surface radius, $J_2$ is a constant
coefficient, and $\theta$ is the angle measured from the planet's
rotation axis.  The gravitational moments, of which $J_2$ is an
example, describe the structure and shape of a rotating fluid and are
measured for the solar system planets by precession rates of satellite and
ring orbits and by spacecraft trajectories.  The gravitational moments
are not known for extrasolar planets, but they generally contribute
small corrections to the gravitational potential. Here, we keep only
the first term of equation~(\ref{eq:gravpot}), the gravitational
potential of a point mass.

The rotational potential $\phi_r$ arises from the force of centripetal
acceleration. For a rigid body\footnote{The gas giant planets
have differential rotation. The slight deviation
from rigid body rotation is due to convective redistribution of
angular momentum.}  rotating at angular velocity $w$ about the $z$
axis,
\begin{equation}
\label{eq:rotpot}
\phi_r =  - \frac{1}{2}w^2 (x^2 + y^2) = -\frac{1}{2}w^2 R(\theta)^2 \sin^2 \theta,
\end{equation}
where the right hand side term is in spherical polar coordinates,
and for axisymmetric rotation $R=R(\theta)$.
Hence the total potential $\phi = \phi_g(\theta) + \phi_r(\theta)$ for
a rigid rotator with a gravitational point mass potential is
\begin{equation}
\label{eq:phitotal}
\phi =  -\frac{GM_P}{R(\theta)} - \frac{1}{2}w^2 R(\theta)^2\sin^2\theta.
\end{equation}
We can use the fact that
the total potential $\phi$ is constant on the surface (Eddington 1926)
to consider that $\phi$ at the pole ($\theta=0$) and at
the equator ($\theta = \pi /2$) are the same. Then equation~(\ref{eq:phitotal})
can be written as,
\begin{equation}
\label{eq:potentialuse}
\frac{GM_P}{R_p} = \frac{GM_P}{R_e} + \frac{1}{2}w^2 R_e^2,
\end{equation}
where the polar radius $R_p$ corresponds to the rotation axis $z$, and
$R_e$ is the equatorial radius.

Then considering the definition of
oblateness, $\eta \equiv (R_e - R_p)/R_p$,
\begin{equation}
\label{eq:omegasq}
\omega^2 = \frac{2GM_P}{R_e^3} \eta,
\end{equation}
and the rotational period $P = 2\pi/\omega$ is
\begin{equation}
\label{eq:period}
P = 2 \pi \sqrt{\frac{R_e^3}{2GM_P}\frac{1}{\eta}}.
\end{equation}
Note that with $e$ as the eccentricity 
(hereafter called ellipticity to avoid confusion with
eccentricity of the planet's orbit) of an ellipse,
$R_p = R_e \sqrt{1 - e^2}$, and so
\begin{equation}
\label{eq:equiv}
\eta \equiv \frac{(R_e- R_p)}{R_p} = \frac{1}{\sqrt{1-e^2}} - 1.
\end{equation}

The maximum oblateness can be estimated by considering the centripetal
acceleration for a given planet mass and radius that can prevent the
body from flying apart, (again, ignoring higher order terms in the
gravitational potential). Considering a particle at $r=R_e$ and
$\theta = \pi/2$,
$GM_Pm/R_e^2 = m \omega^2 R_e$, and considering the relation between
the angular velocity $\omega$ and the oblateness $\eta$
(equation~(\ref{eq:omegasq})), the maximum possible oblateness is
$\eta=0.5$, which corresponds to an ellipticity of $e=0.745$.

It is clear, considering equation~(\ref{eq:period}), that a
measurement of the oblateness $\eta$, together with the measurement of
$M_P$ and $R_P$ (here defined as $R_P = \sqrt{R_e R_p}$)
from the transit photometry and radial velocity data, will yield the
planet's rotational period, subject to two considerations.  The first
is that the $\phi_g$ is expected to deviate from a point mass
potential, since the solar system gas giant planets have non-zero
$J_2$ and higher order moments.  Secondly only the projected
oblateness can be measured and hence only an upper limit on the
rotational period can be obtained. To be precise, suppose the
three-dimensional ellipticity is $e$, the projected ellipticity equals
$e (1 - \cos^2\alpha)$ where $\alpha$ is the angle between the axis of
planet rotation and the line of sight to the planet (see Appendix A of
Hui \& Seager 2002). To the extent the angle $\alpha$ is random, one
expects on average that the projected ellipticity is half the actual
value.  The projected oblateness can be obtained from the projected
ellipticity using the relation between $\eta$ and $e$ in
equation~(\ref{eq:equiv}).  {\it Hereafter, whenever we refer to
ellipticity $e$ or oblateness $\eta$, we mean its projected version,
unless otherwise stated.}

\section{Timescale for Tidal Synchronization of Rotation}
Planetary oblateness is caused by rapid rotation. Planets that rotate
synchronously with their orbital period will generally be rotating too
slowly to be significantly oblate.  Therefore, the tidal evolution
timescale for synchronous rotation will limit the parameter
space---mainly the orbital semi-major axis $D$---where oblate planets
are expected to be found.  Conversely, a detection of oblateness for a
planet together with an estimation of the system's age could help to
constrain the planet's tidal dissipation factor $Q$ (we will refer to
it as $Q_P$ here).  The synchronous rotation timescale is (e.g.
Goldreich \& Soter 1966; Zahn 1977; Hubbard 1984; Guillot et al. 1996)
\begin{equation}
\label{eq:sync}
t_{sync} \approx Q_P \left(\frac{R_P^3}{GM_P}\right) \omega
\left(\frac{M_P}{M_*}\right)^2 \left(\frac{D}{R_P}\right)^6,
\end{equation}
where $\omega = |~\omega_{I} - \omega_{orb}~| \simeq \omega_I$, where
$\omega_{I}$ is the planet's primordial rotational angular velocity
and $\omega_{orb}$ its orbital angular velocity, $R_*$ and $M_*$ are
the star's radius and mass, and $G$ is the universal gravitational
constant. $Q_P$ is inversely related to the tidal dissipation energy.
For a Jupiter-like planet (with Jupiter's tidal $Q_P=10^5$ (Ioannou \&
Lindzen 1993) and Jupiter's current rotation rate $\omega_{I} =
\omega_{Jupiter}= 2\pi/9.92$h) orbiting at 0.05~AU around a solar
twin, $t_{sync} = 2 \times 10^6$ years.  For HD~209458~b, with $R_*
= 1.18 R_{\odot}$, $M_* = 1.06 M_{\odot}$, $R_P = 1.42 R_J$, (Cody \&
Sasselov 2001; Brown et al. 2001), $M_P = 0.69 M_J$ (Mazeh et al. 2000), $t_{sync}=4 \times Q_P$ years.
Because HD~209458~b has $t_{sync}
\ll t_{*}$ (where $t_*$ is the age of the parent star), synchronous
rotation is expected to have been reached. Note that planets of
Jupiter's mass, $Q_P$, and $\omega_I$ orbiting a sun-mass star with $D
\lesssim 0.2$ AU will be synchronous rotators.

The tidal synchronization timescale has as its main
uncertainty $Q_P$ and $\omega_I$.  It is interesting to compare
$t_{sync}$ with the typical ages of stars ($t_*$) with known
extrasolar planets: one to a few billion years.
Figure~\ref{fig:summary}a shows $t_{sync}$ as a function of $D$ and
$Q_P M_P/M_*^2$ to illustrate that planets with $D
\sim 0.15$ to $0.4$~AU have $t_{sync}$ similar to these
typical ages of stars with known planets. Of the parameters in
equation~(\ref{eq:sync}), $D, M_P, R_P$ can be measured for a
transiting planet and $M_*$ can be well estimated.  Therefore $Q_P$
can be constrained with: an oblateness measurement as evidence for
non-synchronous rotation; an assumption for $\omega_{I}$; a planet
with $D$ such that $t_{sync} \sim t_*$. 
(Note that $Q_P$ can also be constrained by considering the
tidal evolution time for tidal circularization, but in this case the
values for $M_P$ and $R_P$ in specific cases are not known.)

In addition to the planet's rotation rate, other orbital parameters
will evolve as a consequence of tides raised on the planet by the
star.  The orbital circularization timescale (Goldreich \& Soter 1966)
and the timescale for co-planarity (defined as the coincidence of the
planet orbital plane and the stellar equatorial plane) are both longer
than the synchronous rotation timescale (Rasio et al. 1966; Hut 1981).
The planet's projected obliquity, $\beta$ (defined here as the {\it
projected} angle of inclination of the planet's rotation axis to the
planet's orbit normal) is measurable for a transiting oblate planet
(see \S\ref{sec-results}) that has an orbital inclination different
from 90$^{\circ}$.  The planet's obliquity is also affected by tidal
evolution.  The timescale for tidal evolution to zero obliquity is the
same as the tidal synchronous rotation timescale for a planet in a
circular orbit (Peale 1999). For eccentric orbits the obliquity
evolution is complex and coupled to other orbital evolution timescales
(Peale 1999). However, a detection of obliquity together with other
measured orbital and physical parameters for a planet in the tidal
evolution regime will be useful to constrain the planet's evolutionary
history. Note that a planet may escape evolving to synchronous
rotation (or to a circular orbit and zero obliquity) in the presence
of other nearby planets. HD~83443 (Mayor et al. 2002) is an
interesting example described in Wu \& Goldreich (2002).

\section{Results and Discussion}
\label{sec-results}
\subsection{Transit Computation}
We want to determine the shape and magnitude of the transit light
curve due to an oblate planet with different values of projected
obliquity. Because the effect is small compared to a spherical planet
transit, the transit light curve must be computed to high accuracy, $
\lesssim 10^{-7}$, in order to accurately calculate effects at the $\sim 10^{-5}$
level.  We use planetocentric coordinates and consider the projected
oblate planet to be an ellipse. The planet's surface is described by
the ellipse equation in polar coordinates,
\begin{equation}
\label{eq:planet}
r^2 = R_e^2 \frac{1-e^2}{(1-e^2 \cos^2(\Theta+\beta - \pi/2))},
\end{equation}
 where $r$ is the distance from planet center and
$\Theta$ is measured from the projected orbit normal.
In this coordinate system, the star's surface is represented by
\begin{equation}
\label{Rstar}
R_*^2 = (r \cos \Theta \pm D \cos i \cos \omega_{orb} (t-t_0))^2 + (r \sin \Theta +  D \sin \omega_{orb} (t-t_0) )^2,
\end{equation}
where $\omega_{orb}$ is the orbital angular velocity, $t$ is time, and
$t_0$ is the time of transit center.  We consider the intersection of
the equations for the planet and star in order to determine (1) start
and end of ingress and egress and (2) integration limits for the star
area blocked by the planet.

\subsection{The Shape and Amplitude of the Oblateness Signature}
The oblateness signature on a transit light curve will be small.  In
order to quantify the shape and magnitude of the oblate planet's
transit signature we compare an oblate planet transit light curve
($T_{ellipse}$) to a spherical planet transit light curve
($T_{sphere}$) for planets of the same projected 
area transiting the same sized
star.  We define $F = [T_{sphere}(t) - T_{ellipse}(t)]$ where $F$ is
the flux difference\footnote{Note that this definition of $F$ is
different from that in Hui \& Seager (2002).} between a spherical
planet (projected to a disk) and an oblate planet (projected to an
ellipse) transit light curve.  The ellipse and projected
sphere are always
chosen to have the same total area i.e. the minimum of $T_{sphere}$
and the minimum of $T_{ellipse}$ are the same in the absence of limb
darkening. Each transit light curve is normalized to the stellar
flux. Figure~\ref{fig:summary}b shows the maximum value of $F$ for
different $e$ and $R_P/R_*$. The parameters $i=90^{\circ}$ and
$\beta=0^{\circ}$ are chosen for this figure---as we will see, this
gives the weakest oblateness signature, and so the contours in Figure
~\ref{fig:summary}b give conservative estimates of the required
photometric precision to detect oblateness.

\begin{figure}
\plotone{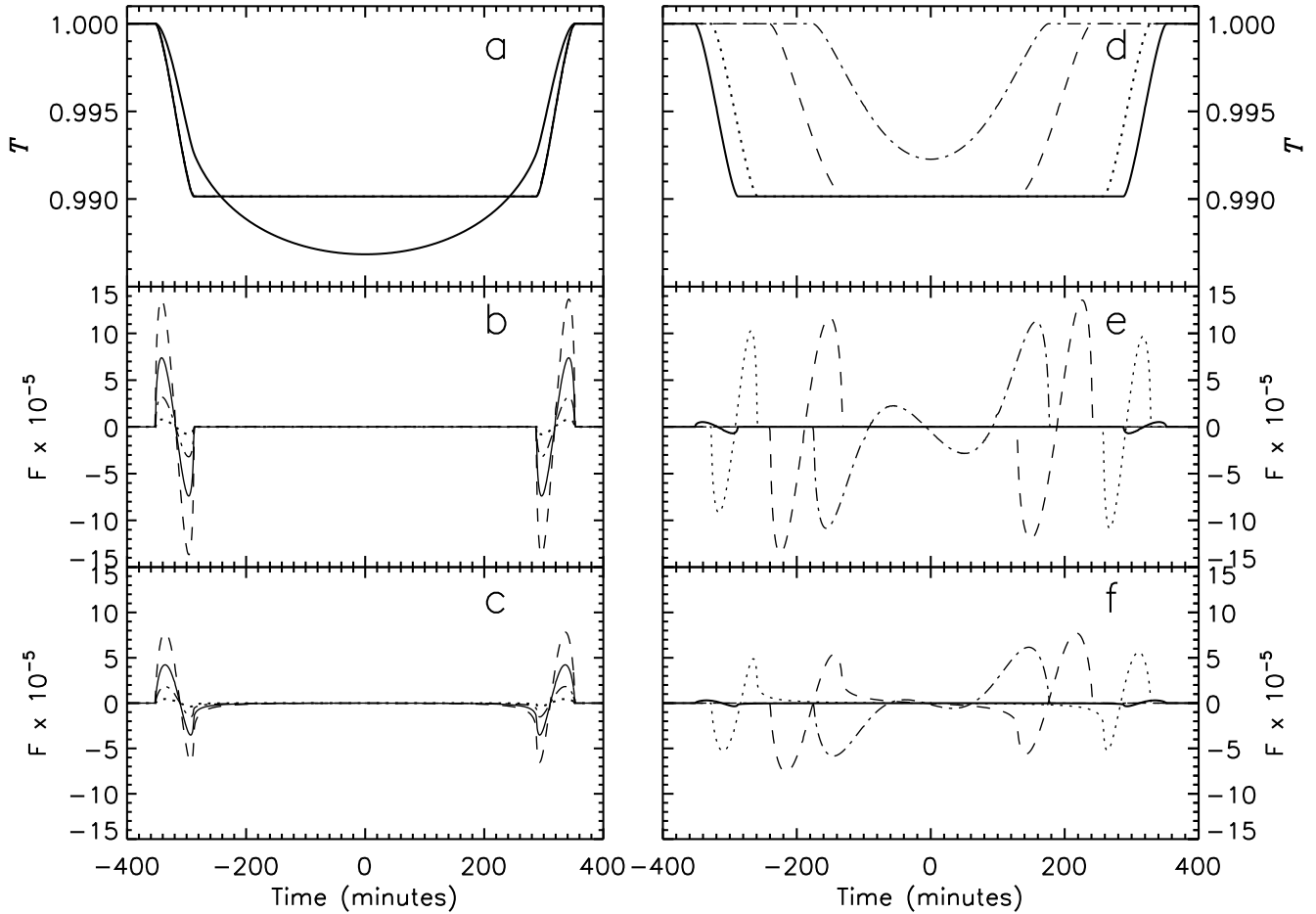}
\caption{A comparison of transit light curves
for spherical and oblate Jupiter-sized planets orbiting a sun-sized
star.  Panel a: The normalized transit light curves ($T$) at orbital
inclination $i=90^{\circ}$---on this scale a transit light curve from
a spherical planet and an oblate planet are indistinguishable. The
round-bottomed curve includes solar limb darkening at 450~nm, whereas
the flat-bottomed curved is computed neglecting limb darkening.  Panel
b: $F = [T_{sphere}(t) - T_{ellipse}(t)]$ (no limb-darkening) where
$T_{ellipse}$ is for planets with projected ellipticity $e=0.1$
(dotted line), $e=0.2$ (short dashed line), $e=0.3$ (solid line) and
$e=0.4$ (long-dashed line).  Panel c: The same $F$ as in panel b, but
at the highly limb-darkened wavelength 450~nm.  Panel d: The
normalized transit light curve ($T$) neglecting limb darkening at
orbital inclinations $i=90^{\circ}$ (solid line), $i=89.47^{\circ}$
(dashed line), $i=88.93^{\circ}$ (dotted line), and $i=88.73^{\circ}$
(dash-dot line). These values of $i$ 
correspond to impact parameters ($=
\cos(i) D / R_*$) 0, 0.4, 0.8, and 0.95 respectively.
Panel e: The flux difference $F = [T_{sphere}(t) - T_{ellipse}(t)]$ at
different $i$ (line styles correspond to $i$ in panel d) for a planet
with $e=0.4$ (that of Saturn) and $\beta=45^{\circ}$.  Panel f: The
same $F$ as in panel e, but for solar limb darkening at 450~nm. The
transit light curves and $F$ in this figure were computed for
$D=0.2$~AU; for other orbital distances the time axis can be scaled by
$(D/0.2{\rm AU})^{1/2}$ (exactly for panels a--c and approximately
for panels d--f).  Note the asymmetry $F(t) \neq F(-t)$ in the transit
light curves in panels e--f (see Figure \ref{fig:asym}).}
\label{fig:ellipse}
\end{figure}

The transit light curve of a Jupiter-sized planet transiting a
sun-sized star with orbital inclination $i=90^{\circ}$ is shown in
Figure~\ref{fig:ellipse}a. The round-bottomed transit is due to limb
darkening, in this case the solar limb-darkening value at 450~nm is
used (Cox 2000). At this blue wavelength the sun is strongly limb
darkened. The flat-bottomed transit was computed neglecting limb
darkening, or in a very non-limb-darkened color (e.g. at IR
wavelengths, although solar limb darkening is minor at $I$ band)
chosen to illustrate the range of possibilities. On the scale in
Figure~\ref{fig:ellipse}a, transits due to oblate and spherical
planets are difficult to tell apart; we show their difference in
subsequent panels.  In Figure~\ref{fig:ellipse} and in this subsection
we consider planets with the same area as Jupiter.  Because Jupiter is
oblate (with $R_e=7.1492 \times 10^{7}{\rm m}$ and $R_p = 6.7137\times
10^{7} {\rm m}$), Jupiter's area corresponds to a sphere with
effective radius $6.9280\times 10^{7}{\rm
m}$. Figure~\ref{fig:ellipse}b shows the flux difference $F =
[T_{sphere}(t) - T_{ellipse}(t)]$ between the non-limb-darkened
transit light curve of a spherical planet and an oblate planet with
ellipticity $e$ varying from 0.1 to 0.4. These transit light curves
were computed for an orbital distance $D=0.2$, but for the case of
$i=90^{\circ}$ the transit time scales as $D^{1/2}$.  The flux
difference $F$ in Figure~\ref{fig:ellipse}b is as high as
1.5$\times10^{-4}$ for an planet with ellipticity $e=0.4$,
corresponding to the oblateness of Saturn.

The oblate planet transit signature can be explained as follows.  At
the start of ingress, when just a small fraction of the planet has
crossed the stellar limb, an oblate planet (projected to an ellipse)
covers less area on the stellar disk than a sphere (projected to a
disk). Hence $F$ (which is actually the flux difference of the light
deficit of a disk and ellipse) is higher.  As ingress progresses, the
area of the star blocked by an oblate planet decreases compared to the
sphere, until the ingress midpoint.  At the midpoint of ingress, when
the planet center coincides with the stellar limb, the area of the
stellar disk that is blocked is roughly the same for an ellipse and a
sphere (except for the deviation of the star limb from a straight
line); hence the fractional difference at mid-ingress is close to
zero. ($F \simeq 0$ at mid-ingress or mid-egress only for
$i=90^{\circ}$ and $R_P \ll R_*$).  The difference $F$ also eventually
reaches around $0$ towards the middle of the transit between ingress
and egress. Another way to consider the oblate transit signature is
that for an oblate planet (with larger $R_e$ than a spherical one of
the same area), the ``first contact'' will occur earlier, and the
``second contact'' will occur later, thus explaining the sign of the
difference signal $F$.

Figure~\ref{fig:ellipse}c shows the same flux differences ($F$s) as in
Figure~\ref{fig:ellipse}b, but at the highly-limb-darkened color of
450~nm. In the presence of strong limb darkening, the maximum value of the
oblateness signature at ingress/egress is lower than in the case of no
limb darkening (Figure~\ref{fig:ellipse}b).  This is because ingress
and egress occur at the stellar limb where a given area takes out less
luminosity than from closer to star center.  Because of limb darkening
the planet transit light curve is asymmetric around mid-ingress. While
this is also true for a spherical planet, the effect is exaggerated
for an ellipse. Limb darkening also causes a flux difference between
an oblate and spherical planet when the planet is fully superimposed
on the parent star; however this difference is negligible compared to
the oblateness signature at ingress and egress.

In all of the above cases, Figure~\ref{fig:ellipse}a-c, $i =
90^{\circ}$ and $\beta = 0$ is assumed.  An oblate planet with $ i <
90^{\circ}$ and non-zero projected obliquity $\beta$ has an
interesting transit light curve.  Figure~\ref{fig:ellipse}d shows the
transit light curve for different orbital inclinations (neglecting
limb darkening).  The dot-dashed line at $i=88.73^{\circ}$ shows an
almost grazing planet transit.  The transit light curves of an oblate
planet and of a spherical planet are indistinguishable on the scale of
Figure~\ref{fig:ellipse}d. Figure~\ref{fig:ellipse}e and
Figure~\ref{fig:ellipse}f show $F$, the flux difference in the transit
light curve between a spherical planet and an oblate planet with
$\beta = 45^{\circ}$ for various orbital inclinations. Note that
ingress and egress last longer for transits with larger $\cos i$,
making the oblateness signature easier to measure.

The ingress and egress oblateness signatures are completely symmetric
about mid-transit (i.e. $F(t) = F(-t)$) for projected
obliquity $\beta=0^{\circ}$ (cases
shown in Figure~\ref{fig:ellipse}c).  The oblateness signature will be
more easily measured from a transit light curve that is asymmetric
about mid-transit in its ingress and egress.  Such an asymmetry will
occur in the case of an oblate planet with a non-zero obliquity and
with orbital inclination different from 90$^{\circ}$.  The asymmetry
is apparent in comparing Figures~\ref{fig:ellipse}b and
\ref{fig:ellipse}e.  Figures~\ref{fig:asym}a and b show the asymmetric
flux difference $F_{asym}= [T_{ellipse}(t) - T_{ellipse}(-t)]$ for
different cases of $i$ and $\beta$.  The oblate planet considered in
Figure~\ref{fig:asym} has $e=0.4$ and the same area as Jupiter.
Reaching amplitudes of $\sim 2 \times 10^{-4}$ these are close to the
current best precision photometry (from HST). The signatures in
Figure~\ref{fig:asym} can be considered maximum signatures using
solar system values. 

In summary, there are {\it two} oblateness signatures one can look
for: 1) a detailed difference in light curves between a transit by a
sphere and a transit by an ellipse, around ingress and
around egress; and 2) an asymmetry of the light curve between ingress
and egress in the case of non-zero projected obliquity $\beta$ and $i
\neq 90^{\circ}$. $\beta$ and $e$ are not degenerate and can both
be extracted from a fit to the light curve.  When both signatures can
be detected, one can obtain constraints on obliquity as well as
oblateness.

We expect to be able to distinguish the oblateness transit signature 
from other similar magnitude effects during transits.
Planetary moons could affect the planet transit light curve but, due
to their size, shape, and projected separation from the planet, would
not mimic planet oblateness (although they could still cause a transit
light curve asymmetry).  Planetary rings will be projected into an
ellipse unless the inclination of the planet's ring plane
to the line of sight is either $0^{\circ}$
or $90^{\circ}$.  With a few parts-per-million photometric precision,
most cases of planetary rings should not be confused with the
oblateness signature because the rings have a relatively large radius
and a slightly different transit signature than an oblate planet. A
planet mass measurement (by radial velocity observations and by the
transit to give $i$) would likely
alleviate any confusion due to transiting
rings (which would make the planet look much larger) due to the
giant planet mass-radius relationship (Guillot et al. 1996). 
Atmospheric lensing (Hui \& Seager 2002) due to a spherical planet may
be confused with the oblateness signature. However at wavelengths
where the planet's atmosphere is strongly absorbing atmospheric
lensing is not present. In addition the transit asymmetry due to
a non-zero $\beta$ is not reproducible by atmospheric lensing.
Confusion at the $10^{-5}$ level from other signals such as
star spots, large planetary features, etc. are unlikely and are
discussed in Hui \& Seager (2002).

\subsection{HD~209458~b: The Only Known Transiting Extrasolar Planet}
Recently Brown et al. (2001) used HST/STIS to measure the transit
light curve of HD~209458~b, the only currently known transiting
extrasolar planet (Charbonneau et al. 2000; Henry et al. 2000). The
resulting photometric precision is at a level of $10^{-4}$. This level
of precision is not high enough to measure the expected oblateness of
HD~209458~b. The close-in extrasolar giant planets like HD~209458~b
are not expected to be significantly oblate due to slow rotational
periods resulting from tidal locking to their orbital
revolution. Assuming synchronous rotation an orbital period of 3.5
days corresponds to an oblateness of 0.0018 ($e=0.05$), considering
HD~209458~b to be a rigid rotator with $M_P = 0.69 M_J$ (Mazeh et
al. 2000) and $R_P=1.42$ (the most recent derivation from Cody \&
Sasselov 2001). At an orbital inclination of 86.1$^{\circ}$ the flux
difference from a transit light curve of a sphere has a maximum
amplitude of only a few $\times 10^{-6}$.  If HD~209458~b has a
non-zero projected obliquity ($\beta$), the asymmetry of ingress to
egress could be as high as 8$\times 10^{-6}$.  Such an obliquity is
not expected, however, due to tidal evolution.  Obtaining an upper
limit for $e$ and $\beta$ for HD~209458~b by a fit to a high-precision
transit light curve would observationally confirm the expected slow
rotational period for HD~209458~b.

\subsection{Constraining the EGP Rotation Rate}
\label{sec-oblatecons}
The four solar system gas giant planets have significant oblateness,
with values listed in Table~1 (note that oblateness here is the actual
oblateness in three dimensions, i.e. not a projected value). With the
exception of Jupiter, they all have substantial obliquity with values
that are consistent with random $\beta$.  The solar system gas giant
planets have measured rotational periods from 9.92~h to 17.24~h (see
Table 1). The rotational periods as calculated from
equation~(\ref{eq:period}) using the measured oblateness values are
within 20\% or less of the actual rotational periods. The estimated
rotational periods are consistently low, due to the neglect of the
higher order gravitational moments in the gravitational potential
(equation~(\ref{eq:gravpot})).

For extrasolar planet transit measurements only the projected
oblateness can be measured, and hence only an upper limit on the
rotational period can be obtained. With a measurement of the rotation
rate together with the projected oblateness $\beta$ could be
constrained. The rotational period could be directly measurable with
the next generation of radio telescopes for
Jupiter-like planets with high synchrotron emission (like Jupiter)
(Bastian, Dulk, \& Leblanc 2000). Note that spectral line measurements
may give information on winds in the planet atmosphere
rather than the planet's rotation rate.

\section{Summary}
\begin{figure}
\plotone{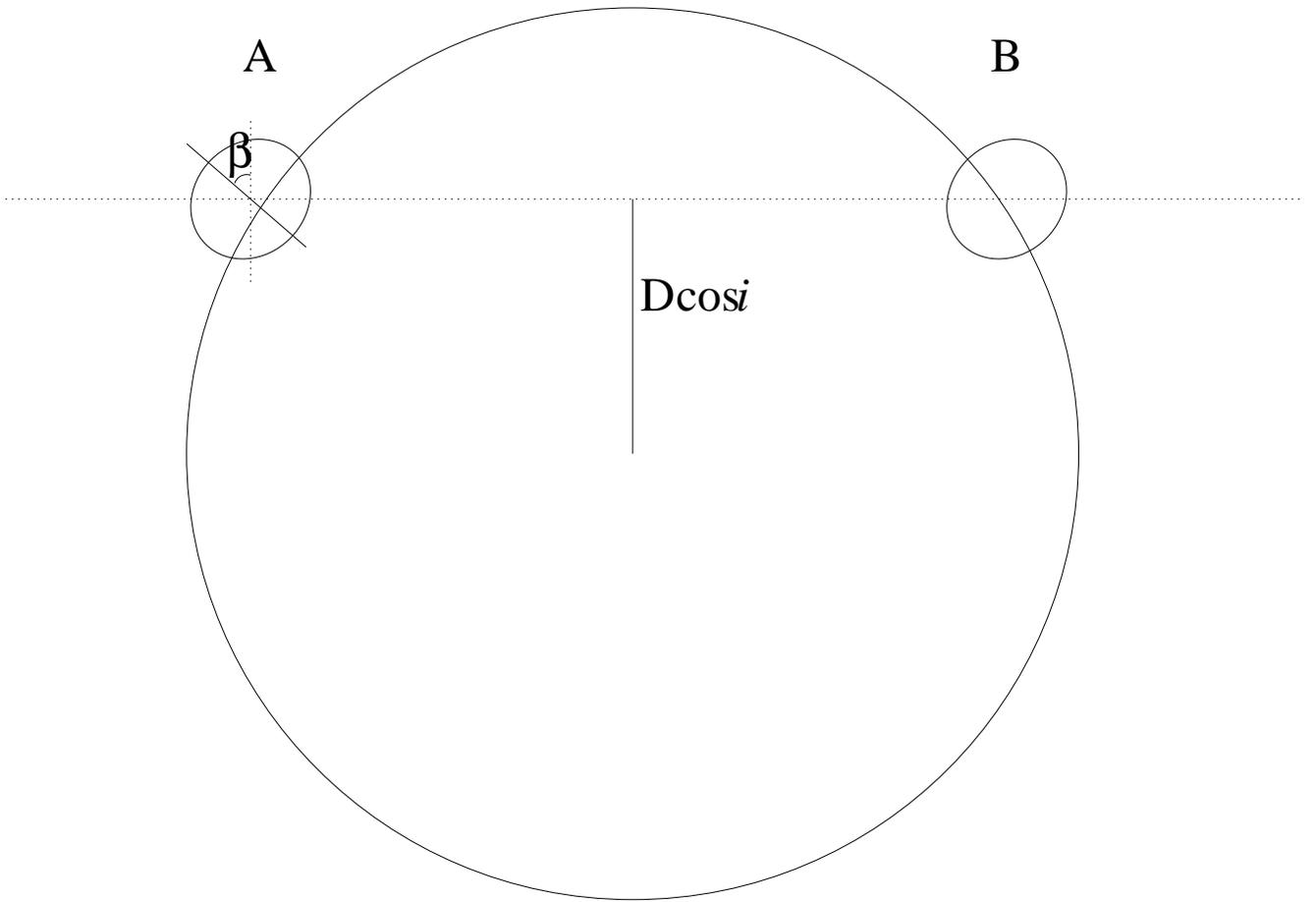}
\caption{Definition of angles.
The projected obliquity, $\beta$, is the projected angle between the
planet's rotation axis and the planet's orbit normal.  The orbital
inclination is $i$ (where $i=90^{\circ}$ corresponds to an edge-on
orbit) and $D$ is the semi-major axis. The planet in this schematic
diagram has an oblateness equal to that of Saturn (10\% which
corresponds to $e=0.4$; note that $e$ in this paper is not orbital
eccentricity but is rather related to oblateness, see \S
\ref{oblatenessperiod}.). For $e>0$, $i < 90^{\circ}$, and $\beta >
0^{\circ}$ the ingress and egress will be asymmetric, for example at
points A and B.}
\label{fig:coord}
\end{figure}

\begin{figure}
\plotone{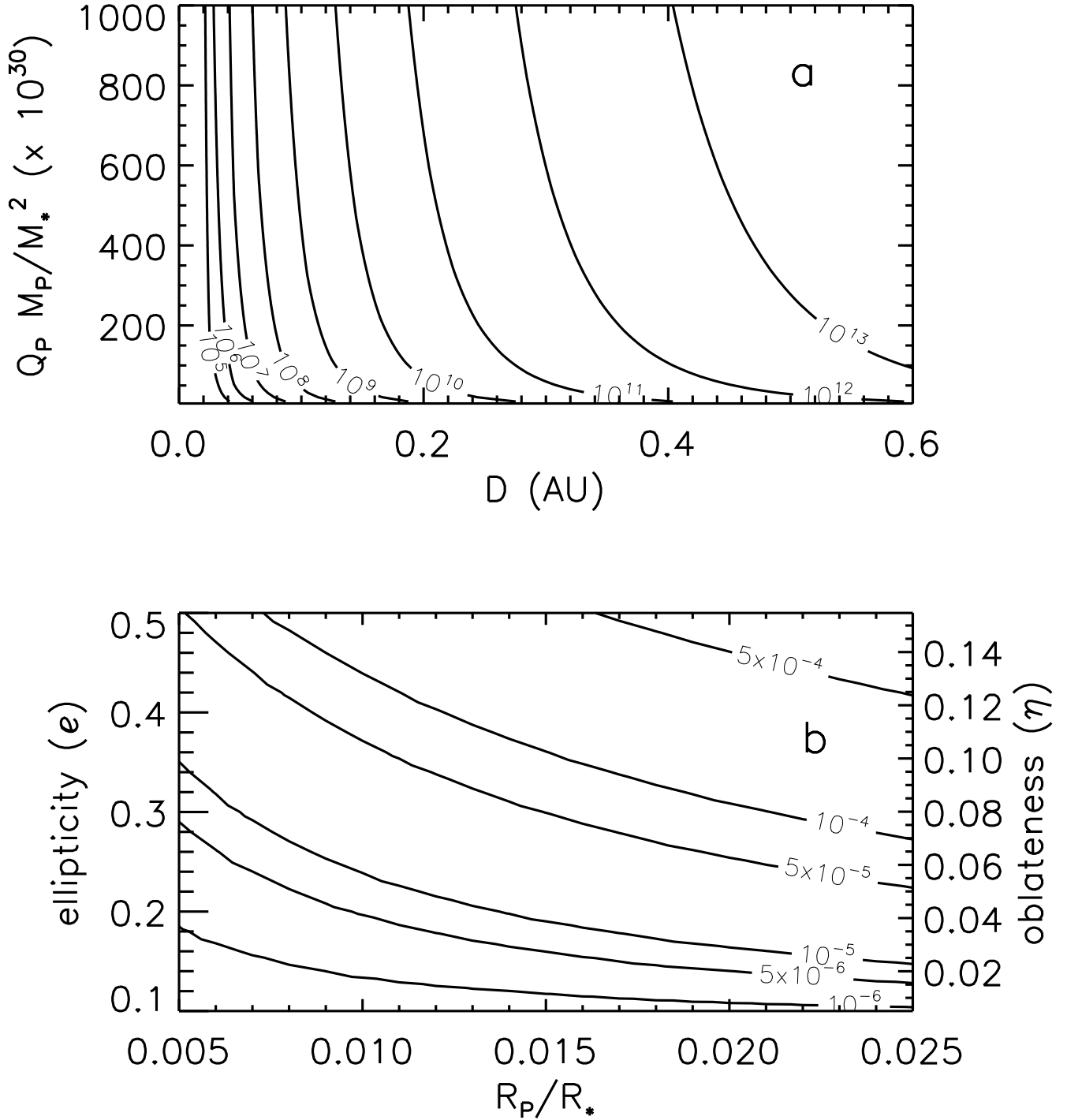}
\caption{
Panel a: Synchronization timescales in years (contour lines) for
different $D$, as a function of $Q_P M_P/M_*^2$.  Typical ages of
stars with known planets are a few billion years; $Q_P$ can be
constrained from an oblateness detection with $t_{sync}$ of the same
timescale. Note that a serious assumption in computing $t_{sync}$ is
that $\omega_{I} = \omega_{Jupiter}=2 \pi /9.92$~h.  Panel b: The
photometric precision (contour lines) required to detect a given
ellipticity as a function of $R_P/R_*$ (neglecting limb darkening).
The parameters $\beta=0^{\circ}$ and $i=90^{\circ}$ were used; because
ellipticity for $\beta >0^{\circ}$ and $i<90^{\circ}$ is much easier
to detect (Figure~\ref{fig:asym}) the required photometric precision
can be considered a conservative estimate.  The corresponding
oblateness, shown on the right $y$ axis, is $\eta =
\frac{1}{\sqrt{1-e^2}}-1$.
}
\label{fig:summary}
\end{figure}

\begin{figure}
\plotone{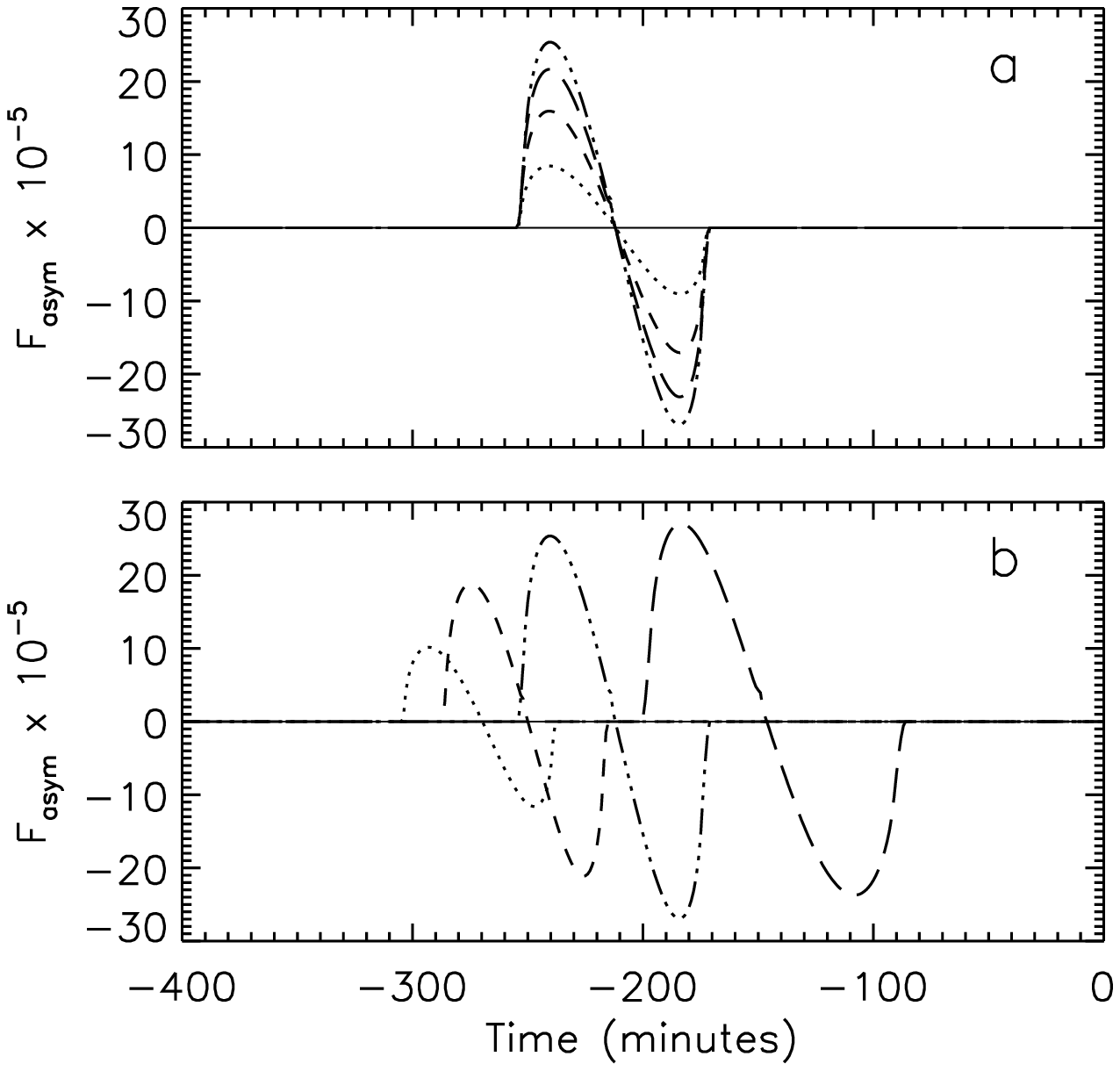}
\caption{The asymmetric flux difference
$F_{asym}= [T_{ellipse}(t) - T_{ellipse}(-t)]$ for a Jupiter-sized
planet transiting a sun-sized star at $D=0.2$~AU versus
time from mid-transit (neglecting
limb darkening).  
Panel a: $F_{asym}$ for $e=0.4$ and
$i=89.2^{\circ}$ (corresponding to an impact parameter of 0.6), with
$\beta=0^{\circ}$ (solid line), $\beta=10^{\circ}$ (dotted line),
$\beta=20^{\circ}$ (short dashed line), $\beta=30^{\circ}$ (long
dashed line), and $\beta=45^{\circ}$ (dot-dash line).  Panel b:
$F_{asym}$ for $e = 0.4$ and $\beta=45^{\circ}$, with $i=90^{\circ}$
(solid line), $i=89.73^{\circ}$ (dotted line), $i=89.47^{\circ}$
(short dashed line), and $i=89.20^{\circ}$ (dot-dash line), and
$i=88.73^{\circ}$ (long dashed line) The $i$ correspond to impact
parameters ($=
\cos(i) D / R_*$) 0, 0.2, 0.4, 0.6, and 0.8, respectively.
Note the solid lines in each panel
show that there is no transit asymmetry for an
oblate planet with $\beta=0^{\circ}$ or $i=90^{\circ}$.}
\label{fig:asym}
\end{figure}

An oblate planet and a spherical planet of Jupiter's area orbiting a
sun-like star will have different ingress and egress transit light
curves by as much as a few to 15 $\times 10^{-5}$.  A planet's
oblateness is most easily detectable in the case where the planet's
orbital inclination is different from edge-on, and the planet's
projected obliquity is substantially different from zero. In this case
the transit light curve is asymmetric between the ingress and egress
(see Figure~\ref{fig:coord} and Figure~\ref{fig:asym}), a signature
that can be detected by folding the light curve around mid-transit. The
oblateness signature due to a planet with zero projected obliquity is
symmetric in ingress and egress; the oblateness can then be measured
using a model fit to the data (by exploiting the detailed differences
in the light curves produced by a sphere and an ellipse; see
Figure~\ref{fig:summary}).  The Kepler mission is expected to find 30
to 40 transiting giant planets with orbital semi-major axes $< 1$~AU,
about 20 of which will be at $D > 0.2$~AU, and Kepler and three other
space telescopes will reach part-per-million photometric precision.
The future is promising for the oblateness measurement of extrasolar
planets.

\acknowledgements{We thank Scott Gaudi for reading the manuscript,
and Jordi Miralda-Escude, Scott Gaudi, Gabriela Mallen-Ornelas, and
Eliot Quataert for useful discussions. We thank the referee Bob Noyes
for a careful reading of the manuscript. S.S. thanks John Bahcall for
valuable advice and support.  S.S. is supported by the W.M. Keck
Foundation. L.H. is supported by an Outstanding Junior Investigator
Award from the DOE and grant AST-0098437 from the NSF.}

\begin{table}
\begin{tabular}{l l l l l l l}
\hline
Planet & Oblateness & $e$ & period (h) & period (h) & obliquity \\ & $\eta \equiv \frac{R_e - R_p}{R_p}$ & &(measured)
 &(calculated) &  \\ \hline
\\ \hline
Jupiter & 0.0648744 & 0.34  & 9.92  & 8.23 &3.12 \\
Saturn & 0.0979624  & 0.41 & 10.66 & 9.47  & 26.73 \\
Uranus & 0.0229273  & 0.21 & 17.24 & 13.83 & 97.86 \\
Neptune& 0.0171      & 0.18 & 16.11 & 14.07 & 29.58 \\
\\ \hline
\\
\end{tabular}
\caption{Solar system giant planet oblateness, period,
and obliquity from Cox (2000). Also listed is the period calculated
from the oblateness by equation~(\ref{eq:period}) (using $M_P$ and
$R_e$ from Cox (2000)).}
\end{table}

\begin{references}
\reference{} Bastian, T. S., Dulk, G. A., \& Leblanc, Y. 2000, ApJ, 545, 1058

\reference{} Brown, T. M., Charbonneau, D., Gilliland, R. L., Noyes, R. W., Burrows, A. 2001, ApJ, 552, 699

\reference{} Chandrasekhar, S. 1969, Ellipsoidal Figures of Equilibrium, Yale University Press

\reference{} Charbonneau, D., Brown, T. M., Latham, D. W., Mayor, M.
2000, ApJ, 529, L45

\reference{} Charbonneau, D., Brown, T. M., Noyes, R. W., \& Gilliland, R. L. 2002, ApJ, 568, 377

\reference{} Collins, G. W., II. 1963, ApJ, 138, 1134

\reference{} Cody, A., \& Sasselov, D. D. 2001, submitted to ApJ, astro-ph/0111494

\reference{} Cox, A. N. 2000, Allen's Astrophysical Quantities,
(New York: Springer-Verlag)

\reference{} Danby, J. M. A. 1962, Fundamentals of Celestial Mechanics, (New York: Macmillan)

\reference{} Eddington, A. 1926, The Internal Constitutions of the Stars,
(New York: Dover)

\reference{} Goldreich, P., \& Soter S. 1966, Icarus, 5, 375

\reference{} Guillot, T., Burrows, A., Hubbard, W. B., Lunine, J. I., \& Saumon, D. 1996, ApJL, 459, 35

\reference{} Hubbard, W. B. 1984, Planetary Interiors, Van Nostrand Reinhold Co., New York

\reference{} Hui, L., \& Seager, S. 2002, ApJ, in press, astro-ph/0103329

\reference{} Hut, P. 1981, A\&A, 99, 126

\reference{} Ioannou, P. J., \& Lindzen, R. S. 1993, ApJ, 406, 266

\reference{} Mayor, M., Naef, D., Pepe, F., Queloz, D., Santos, N.,
Udry, S., \& Burnet, M. 2002, in Planetary Systems in the Universe:
          Observation, Formation and Evolution, IAU Symp. 202, Eds.
          A. Penny, P. Artymowicz, A.-M. Lagrange and S. Russel ASP
          Conf. Ser, in press

\reference{} Mazeh, T. et al. 2000, ApJ, 532, L55

\reference{} Peale, S. J. 1999, ARAA, 37, 533

\reference{} Rasio, F. A., Tout, C. A., Lubow, S. H., \& Livio, M. 1996, ApJ, 470, 1187

\reference{} Seager, S., Whitney, B. A., \& Sasselov 2000, ApJ, 540, 504

\reference{} van Belle, G. T., Ciardi, D. R., Thompson, R. R.,
Akeson, R. L., Lada, E. A. 2001, ApJ, 559, 1155

\reference{} Wu, Y., \& Goldreich, P. 2002, ApJ, 564, 1024

\reference{} Zahn, J. P. 1977, A\&A, 57, 383

\end{references}
\end{document}